\documentclass[12pt]{article}
\usepackage{times}
\usepackage{geometry}
\usepackage[utf8]{inputenc}
\usepackage{enumitem,amssymb}
\usepackage{ragged2e}
\usepackage{graphicx}
\usepackage{comment}
\usepackage{multicol}
\usepackage[usenames]{xcolor} 
\definecolor{xlinkcolor}{cmyk}{1,1,0,0}

\usepackage[
 colorlinks=true,    
 linkcolor=xlinkcolor,     
 citecolor=xlinkcolor,     
 filecolor=xlinkcolor,  
 urlcolor=xlinkcolor,      
 final=true
]{hyperref}

\usepackage[numbers,sort&compress]{natbib}
\setenumerate{itemsep=0mm}
\usepackage{indentfirst}

\usepackage{dsfont}
\usepackage{amsmath}
\numberwithin{equation}{section}

\IfFileExists{dsfont.sty}
	{\usepackage{dsfont}
         \let\mathbb=\mathds
         \newcommand{\id}{\mathds{1}}}
	{\typeout{Package dsfont.sty was not found, using alternative macros.}
         \let\mathds=\mathbb
         \newcommand{\id}{\mbox{1 \kern-.59em \textrm{l}}}}
         \newcommand{\sdfrac}[2]{\mbox{\small$\displaystyle\frac{#1}{#2}$}}
         
\usepackage{tikz}
\usepackage{lipsum}
\usepackage{microtype}
\usepackage{fancyhdr}
\usepackage{gensymb}
\usepackage{wrapfig}

\usepackage{titlesec}
\titleformat{\section}
  {\normalfont\bfseries}{\thesection}{1em}{}
\titleformat{\subsection}
  {\normalfont\normalsize\itshape}{\thesubsection}{1em}{}
\titlespacing*{\section}
{0pt}{6pt}{4pt}
\titlespacing*{\subsection}
{0pt}{4pt}{2pt}

\renewcommand{\a}{\alpha}
\renewcommand{\b}{\beta}
\newcommand{\g}{\gamma}
\renewcommand{\d}{\delta}
\newcommand{\e}{\epsilon}
\newcommand{\ve}{\varepsilon}

\renewcommand{\k}{\kappa}
\newcommand\vk{\varkappa}
\renewcommand{\l}{\lambda}
\newcommand{\m}{\mu}
\newcommand{\n}{\nu}

\newcommand\p{\pi}
\renewcommand{\r}{\rho}
\newcommand{\s}{\sigma}
\renewcommand{\t}{\tau}

\newcommand{\vf}{\varphi}

\newcommand{\w}{\omega}

\newcommand\D{\Delta}
\renewcommand{\L}{\Lambda}
\newcommand{\rH}{r_{\!_H}}
\newcommand{\GN} {G_{\scriptstyle N}}

\newcommand{\be}{\begin{equation}}
\newcommand{\ee}{\end{equation}}
\newcommand{\bea}{\begin{eqnarray}}
\newcommand{\eea}{\end{eqnarray}}
\newcommand{\bes}{\begin{subequations}}
\newcommand{\ees}{\end{subequations}}

\newcommand{\pa}{\partial}
\newcommand{\cA}{{\cal A}}
\newcommand{\cL}{{\cal L}}
\newcommand{\tF}{\tilde F}
\def\lag{\langle}
\def\rag{\rangle}
\newcommand{\na}{\nabla}
\newcommand{\bx}{{\bf x}}
\newcommand{\br}{{\bf r}}

\def\nbox#1#2{\vcenter{\hrule \hbox{\vrule height#2in
\kern#1in \vrule} \hrule}}
\def\sq{\,\raise2pt\hbox{$\nbox{.10}{.10}$}\,}

\newcommand{\sumi}{\raisebox{-1.5ex}{$\stackrel{\textstyle\sum}{\scriptstyle i}$}}
\begin{document}

\begin{centering}
{\LARGE
Snowmass2022: TF01 White Paper}\\[+1em]

{\fontfamily{phv}\selectfont{\Large\bfseries Beyond Einstein's Horizon: \\
Gravitational Condensates and Black Hole Interiors\\[+9pt]
in the Effective Theory of Gravity}}\\[+1.5em]

{\large
Emil Mottola \\
Dept. of Physics and Astronomy\\
Univ. of New Mexico \\
Albuquerque NM 87131\\
\vspace{2mm}
{\it E-mail:}\ {\fontfamily{lmtt}\selectfont \href{mailto:emottola@unm.edu}{emottola@unm.edu},\,  
\href{mailto:mottola.emil@gmail.com}{mottola.emil@gmail.com} }\\[+3em]
Abstract  \\[+1.5em]}

\end{centering}


Two of the most fundamental problems at the nexus of Einstein's classical General Relativity (GR) and Quantum Field Theory (QFT) are:
\vspace{-3mm}
\begin{enumerate}[label=(\arabic*)]
\item complete gravitational collapse, presumed in classical GR to lead to a Black Hole (BH) horizon and interior singularity, which 
generate a number of paradoxes for quantum theory; \\[-3ex]
\item the origin and magnitude of the cosmological dark energy driving the accelerated expansion of the Universe. 
\end{enumerate}

\vspace{-2mm}
In this Snowmass white paper it is proposed that these twin puzzles on disparate scales are related, and that their resolution 
depends upon taking full account of the conformal anomaly of quantum matter in gravitational fields. The topological term in 
the anomaly leads naturally to the introduction of an abelian $3$-form gauge field, whose field strength can account for a 
variable gravitational condensate with the vacuum dark energy equation of state $p=-\r$, the magnitude of which depends 
upon macroscopic boundary conditions rather than ultraviolet cutoffs. The resulting Effective Field Theory (EFT) of low energy 
quantum gravity results in a non-singular `BH' interior and physical surface replacing the classical event horizon, which is a 
gravitational condensate star free of any information paradox. The development and predictions of this EFT can be tested by 
gravitational waves and observational cosmology in the coming decade.

\clearpage

\pagestyle{plain}
\pagenumbering{arabic}

\section{Introduction: Two Problems of Some Gravity}

At the most microscopic scales so far probed, the principles of Quantum Field Theory (QFT) hold, and matter under extreme pressures 
and densities is clearly quantum in nature. Yet Einstein's classical General Relativity (GR) remains unreconciled to quantum theory, 
and tests of GR are still based on an essentially classical description of matter and energy. This tension between quantum matter 
and classical gravity comes to the fore in the twin puzzles and paradoxes of Black Holes (BHs) and vacuum dark energy.

\subsection{The Puzzles and Paradoxes of Black Holes}

Classical BHs are vacuum solutions of Einstein's equations, in which all details of the quantum matter that gave rise to them are 
subsumed into an interior spacetime singularity of infinite pressure and density. According to the classical singularity
theorems, as long as a closed trapped surface forms and the collapsing matter satisfies certain energy conditions, a BH singularity is 
unavoidable~\cite{Penrose:1965,HawkPen:1970}. In the case of rotating BHs, analytic extension of the exterior Kerr solution 
through the horizon leads not only to singularities but also to closed timelike curves in their interior~\cite{HawkEllis:1973}, 
violating causality at even macroscopic scales far larger than the microscopic Planck scale. 

When quantum effects are considered, additional problems appear at the macroscopic horizon scale. Since a BH horizon is a 
marginally trapped surface from which no matter or information can escape, at least classically, the entropy of matter falling into
a BH would seem to have vanished from the external universe. This raises potential conflicts both with statistical physics and
quantum unitary evolution. Bekenstein's suggestion that the BH horizon itself be assigned an entropy proportional to its area
$A_{_H}$ provides one possible route to salvaging the second law of thermodynamics~\cite{Beken:1973}. This proposal
received some support when Hawking found that a non-rotating BH of mass $M$ could radiate thermally at the Hawking 
temperature~\cite{Hawk:1975,HartHawk:1976}
\vspace{-1mm}
\be
T_{_H} = \frac{\hbar c^3}{8\pi k_{_B} GM} \simeq 6 \times 10^{-8}\,^\circ \!K \left(\frac{M_\odot\!}{\! M}\right)
\label{TH}
\ee
\vspace{-1mm}
in solar units. The variation of the {\it classical} mass-energy of the BH can then be written~\cite{Smarr:1973}  
\be
dE = dM c^2 = \frac{c^2}{8 \pi G}\,  \k_{_H} dA_{_H}  
= \frac{\hbar c^3}{8\pi k_{_B} GM}\ d\left(\frac{4\pi k_{_B}GM^2}{\hbar c}\right) =T_{_H} d S_{_{BH}}
\label{dEdS}
\ee
in which $\hbar$ cancels out, but suggestive of the first law of thermodynamics, with $\k_{_H}\!=\! c^2/2\rH$ the 
surface gravity of a Schwarzschild BH, and the Bekenstein BH entropy $S_{_{BH}}$ identified as
\be
S_{_{BH}} =  \frac{4\pi k_{_B}GM^2}{\hbar c}= \frac{A_{_H}}{4L_{Pl}^2\!} \simeq 10^{77}\, k_{_B} \left(\frac{\! M}{M_{\odot}}\right)^2
\label{SBH}
\ee
or one quarter the horizon area $A_{_H}\!=\! 4 \pi \rH^2$ in Planck units, where $L_{Pl}\!= \!\!\sqrt{\hbar \GN/c^3}\! \simeq\!1.6 \times\! 10^{-33}$cm.

According to the thermodynamic interpretation of (\ref{TH})-(\ref{SBH}), an arbitrarily heavy BH would have an arbitrarily 
low temperature, but an arbitrarily high entropy. This is completely unlike any other ultracold quantum system, which one would expect
should settle into its quantum ground state with vanishingly {\it small} entropy as $T\!\to \!0$. The enormous value of $S_{_{BH}}$ 
in (\ref{SBH}) is also some $19$ orders of magnitude larger than the entropy of a stellar progenitor of the mass of the sun. 
It is not clear how this is consistent with Boltzmann's formula $S = k_{_B}\!\ln W(E)$ relating the entropy of a system to 
its total number of microstates $W(E)$ at fixed energy $E$, or how the necessary $\exp(10^{19})$ additional microstates 
could be associated with the horizon, which is assumed to be a mathematical boundary only, with no special local dynamics
and a negligibly small local curvature for $M$ large.

The inverse relation between the mass of a BH and its temperature, (\ref{TH}), results also in a negative heat capacity 
$dE/dT < 0$~\cite{HawkSpecHeat:1976} which is very large in magnitude, implying a highly unstable situation, 
inconsistent with a stable ground state and (near) equilibrium thermodynamics. This negative heat capacity of a Schwarzschild
BH is in marked contrast with extremal BHs with angular momentum $J \!=\! GM^2/c$, for which the Hawking temperature 
vanishes and which may be related by duality arguments to supersymmetric theories that have a stable ground 
state~\cite{StromVafa:1996}.

It has also been noted that although the Hawking temperature $T_{_H}$ in (\ref{TH}) far from the BH 
is very low, the {\it local} temperature of the radiation\\
\vspace{-5mm}
\begin{wrapfigure}{r}{.5\textwidth}
\vspace{-6mm}
\be
T_{loc}(r) =\ \,\frac{T_{_H}}{\hspace{-3mm}\sqrt{\raisebox{-1pt}{$1\ $--}\ \sdfrac{\raisebox{-1.5ex}{$r_H$\!}}{r}}}
\label{Tol}
\vspace{-5mm}
\ee
\end{wrapfigure}
is arbitrarily high when extrapolated back to the vicinity of the horizon $r \!\to\! \rH$ by the Tolman redshift relation~\cite{Tolman:1930},
even becoming transplanckian in this limit. It is not clear why thermal fluctuations at transplanckian temperatures should not have 
significant backreaction effects on the classical spacetime geometry near the horizon, or equivalently, why it should be valid to
assume a rigidly fixed classical BH background, and neglect quantum fluctuations in the stress tensor down to $r\!=\!\rH$, as
the original Hawking argument assumes.

Lastly, the thermal radiation emerging from the BH in Hawking's original treatment is in a mixed thermal state, 
which would imply a breakdown of quantum unitary evolution, that is difficult, if not impossible to recover at the late 
or final stages of the Hawking BH evaporation process~\cite{HawkUnit:1976,Preskill:1992,GiddNelson:1992,Page:1993,UnruhWald:2017}.

These myriad BH puzzles and paradoxes are so serious that it has been suggested their resolution may require the abandonment 
of the basic physical principles of GR or QFT~\cite{Preskill:1992,Mathur:2009,AMPS:2013,Giddings:2013,MottVauPT}, or perhaps point 
the way to a more fundamental theory replacing both. This accounts for why BHs have remained an active area of fundamental research
for more than four decades, with many interesting proposals up to the present~\cite{AHMST:2020}. Yet the solution remains elusive,
and may require a radical revision of the classical view that nothing in particular happens at the horizon~\cite{Giddings:2017}.
The `fuzzball' proposal postulates an enormous number of macroscopic stringy states on the scale of the BH horizon~\cite{Mathur:2015},
which may not admit a smooth average classical metric description of the BH interior at all. Quantum scrambling of information 
may radically change the semi-classical picture of BH evaporation and interior~\cite{HarlowBHs:2016}. The `firewall' proposal also 
seems to imply some quantum effects becoming important on the horizon as well~\cite{AMPS:2013}, in apparent conflict with the 
usual semi-classical or Effective Field Theory (EFT) approaches to macroscopic gravity, that rely on an expansion in local curvature
invariants~\cite{Don:1994,Bur:2004}, and imply only negligibly small effects for large BHs. 

\subsection{The Puzzle of Cosmological Vacuum Dark Energy}

On the completely different scale of cosmology, observations of type Ia Supernovae (SNe) at moderately large redshifts lead 
to the conclusion that the expansion of the Universe is accelerating \cite{RiessSN:1998,PerlmutterSN:1999}. This is possible 
in classical GR only if the dominant energy in the Universe has an effective equation of state with $\r + 3p <0$, {\it i.e.} assuming 
its energy density $\r >0$, it must have {\it negative} pressure. Taken at face value, the SNe observations imply that some 
$\sim 70\%$ of the energy in the Universe is composed of a $p\!=\! -\r$ `fluid,' interpreted as a positive cosmological constant of~\cite{DES:2019}
\vspace{-1mm}
\be
\L_{SN}\simeq  3.2 \times 10^{-122}/L_{Pl}^2\,.
\label{LSN}
\vspace{-1mm}
\ee

From time of Pauli it has been recognized that the $\L$ term has the same form as the energy density of the vacuum in
QFT, in which it appears as a naively quartically divergent quantity. When compared to the Planck scale and $\GN$, the value 
of $\L_{SN}$ in (\ref{LSN}) represents the most severe scale hierarchy problem in all of physics, clashing with
expectations of `naturalness' developed over several decades in the Effective Field Theory (EFT) context, as well as the
reasonable expectation that macroscopic gravity and the value of $\L$ at cosmological scales should not require a UV complete 
quantum theory valid down to the microscopic Planck scale. Cosmological dark energy thus challenges our notions of
EFT applied to gravity in some ways similar to the BH puzzle and information paradox. The value of $\L_{SN}$ 
also resembles the hierarchy problem of the Higgs mass in the Standard Model (SM), which suggests that scale or
conformal invariance may play a role.

Standard $\L$CDM Cosmology also assumes that the present slight ($\sim\!\! 10^{-5}$) anisotropies in the Cosmic Microwave Background
were generated during a very early epoch of cosmic inflation, dominated by a much larger {\it effective} $\L_{\rm eff}$ vacuum energy of a 
scalar inflaton field, which later somehow `relaxed' to its present, unexplained value. Quantum fluctuations of the inflaton are then
invoked to give rise to the CMB anisotropies observed today. Their nearly scale invariant Harrison-Zel'dovich spectrum is a consequence 
of (approximate) de Sitter symmetry \cite{CMB_AntMazEM:1997,AntMazEM:2012}. On the other hand de Sitter space spontaneously
creates particle pairs from the vacuum \cite{EMdS:1985,AndEM:2014a,AndEM:2014b}, leading to de Sitter vacuum instability in 
QFT~\cite{Polyakov:2008,AndEM:2018}, as well as infrared divergent/de Sitter breaking 
propagators~\cite{AntEM_dSJMP:1991,MorTsamWood:2012} and interactions~\cite{Polyakov:2008}. De Sitter space is also 
problematic in string theory~\cite{DanRiet:2018}. These theoretical results together with observational cosmology
suggest that the de Sitter invariant `vacuum' with any constant positive value of $\L$ is not a stable quantum ground state, and 
instead that quantum processes themselves could lead to dynamically changing vacuum energy, 
{\it cf.}~\cite{VacDecay_FAFEM:1987,AntMazEM:2007}.  

The order of magnitude of (\ref{LSN}), no less than inflationary models, the CMB and the problems with de Sitter space together 
suggest that a critical element of the EFT of macroscopic gravity is missing, which would relate $\L_{\rm eff}$ to the cosmological 
Hubble scale rather than to the microscopic Planck scale. In that case the dimensionless number to be explained by the EFT would
become $0.70$ rather than of order $10^{-122}$. Since the Hubble scale is a dynamical quantity, dependent upon the evolution 
and age of the Universe, such a relation would require $\L_{\rm eff}$ to be dynamical as well. 

The purpose of this contribution is to sketch a proposal for the general EFT of low energy gravity in which dynamical 
vacuum energy is deduced from general principles of QFT and the conformal anomaly, with one additional input of topological 
vacuum susceptibility suggested by the analogy to the chiral suspeceptibity of QCD. While the observational
predictions of this proposed EFT of gravity remain to be fully worked out, all the elements appear to be in place to directly address 
and potentially resolve both the puzzles and paradoxes of BHs and cosmological vacuum dark energy, within the same  
fundamental theoretical framework based on first principles.

This Snowmass white paper contribution contains the sketch of the main ideas and elements of the EFT, and a proposed program for
a wider physics community. Additional details of the EFT are presented in arXiv:2205.04703~\cite{EMEFT:2022}.

\section{Conformal Anomaly Effective Action}
\label{Sec:Anom}

The first essential element in constructing the low energy EFT of gravity relevant to both BHs and cosmological
dark energy is the conformal anomaly in the trace of the stress-energy tensor of QFT, and the effective action corresponding to it.  

In GR the stress-energy source $T^{\a\b}$ for Einstein's eqs.~is treated as completely classical, whereas in QFT 
$\hat T^{\a\b}$ is an operator. Since SM matter is certainly quantum in nature, replacing the quantum operator $\hat T^{\a\b}$ 
by its  (renormalized) expectation value $\lag \hat T^{\a\b}\rag$ in a semi-classical approximation amounts to assuming that 
$\hat T^{\a\b}$ is infinitely sharply peaked at its mean value. However, it is readily verified that  {\it e.g.} $\big\lag \hat T^{\a\b} (x)\, \hat T^{\m\n}(y)\big\rag - \big\lag \hat T^{\a\b}(x) \big\rag\, \big\lag \hat T^{\m\n}(y)\big\rag \neq 0$ at one-loop order. 
Connected correlation functions of this kind probe the polarization and entanglement properties of the quantum vacuum 
due to gravity, and possess short distance singularities as $(x-y)^2\! \to\! 0$. In semi-classical gravity, simple dimensional 
analysis as well as explicit calculations of $\lag TT\rag$ correlation functions show that if the spacetime metric varies on the length 
scale $\ell$, then quantum effects are of parametrically of order $L_{Pl}^2/\ell^2$ relative to $\lag T^{\a\b}\rag$.  Since the
corresponding energy scale of $M_{Pl} c^2 =1.2 \times 10^{19}$ GeV is so much higher than those probed in either accelerator
experiments or astrophysics, it is generally assumed that quantum effects in macroscopic gravity can be neglected. 

There is however a second kinematic regime where the Lorentz invariant distance $(x-y)^2 \to 0$, and where two and higher
point quantum correlation functions are singular, namely on the light cone. Since light cones extend over arbitrarily large distances, 
these lightlike correlations are not limited to the ultrashort $L_{Pl}$, but can lead to {\it macroscopic} quantum effects.
The two very different sorts of quantum effects, ultraviolet {\it vs.} lightlike, can be distinguished by the behavior of connected
QFT correlators in momentum space, and require two quite different EFT treatments.

Short distance UV quantum corrections to GR can be taken into account by adding to the Einstein-Hilbert action of classical GR,
terms higher order local curvature invariants, such as $R_{\a\b\m\n}R^{\a\b\m\n}, R_{\a\b}R^{\a\b}, R^2, \sq R$. These terms 
are dimension four operators $\sim (\pa^2 h_{\a\b})^2$ in terms of the metric degrees of freedom $h_{\a\b}$, in contrast to 
the dimension two Einstein-Hilbert (E-H) action. This is consistent with the most common EFT approach, pioneered by Wilson and Weinberg 
of taking short distance degrees of freedom into account by an expansion in ascending powers of higher derivatives of the local
invariants, divided by appropriate powers of the UV cutoff scale, expected to be the Planck scale $M_{Pl}$ for gravity. The basis 
of this approach and suppression of higher derivative local terms in the effective action is the decoupling of heavy degrees of freedom. 

On the other hand it has also been known for some time that QFT anomalies are not captured by such an expansion in higher 
order local invariants, or suppressed by a UV cutoff scale. Anomalies are associated instead with the fluctuations 
of massless fields which do not decouple, and which are characterized by $1/k^2$ poles in momentum space correlation functions, 
that grow large on the light cone $k^2\! \to\! 0$ rather than the extreme UV regime $k^2 \sim M_{Pl}^2$. In $3+1$ dimensions 
massless poles have been found by explicit calculation in the triangle anomaly diagrams of 
$\lag J_5^\l J^\a J^\b\rag, \lag T^{\m\n} J^\a J^\b\rag$ in massless QED$_4$~\cite{GiaEM:2009}, and most recently in the 
$\lag T^{\a\b} T^{\g\l} T^{\m\n}\rag$ of a general conformal field theory (CFT),  by solution of the conformal Ward Identities in momentum space~\cite{TTTCFT:2019}.

The anomaly pole implies the propagation of at least one additional light ({\it a priori} massless) degree of freedom 
in the low-energy EFT, not apparent from the classical Lagrangian. This additional degree of freedom is a collective mode of the 
underlying many-particle QFT~\cite{GiaEM:2009,EMZak:2010,TTTCFT:2019}. The effective action of the anomaly is correspondingly 
non-local in terms of the original fields, explaining why it is not captured by an expansion in powers of local gauge fields in QCD or
curvature invariants in gravity. The light cone singularity of anomalous correlators in momentum space is the signature of the 
low energy properties of the anomaly, responsible for {\it e.g.} $\p^0\!\to\!2\g$,  distinct from the usual short distance singularity 
of quantum correlation functions. The effective action of the anomaly can be rendered into a local form only by the introduction 
of at least one new local field in the EFT which transforms non-trivially under the action of the anomalous symmetry. 

In $D\!=\!4$ the mean value of $\lag \hat T^{\a\b}\rag$, renormalized by any method that preserves its covariant 
conservation also results in it acquiring an anomalous trace in background gravitational and gauge fields, the general form 
of which is~\cite{CapperDuff:1974,Duff:1977,BirDav}
\vspace{-1.5mm}
\be
\big\lag \hat T^\a_{\ \ \a} \big\rag\ =  \ b\, C^2 + b'\, \left( E - \tfrac{2}{3} \sq R\right) + \sumi \ \b_i\, \cL_i \equiv \ \frac{\cA}{\!\!\!\sqrt{-g}}
\label{tranom}
\vspace{-1mm}
\ee
even if the underlying QFT is otherwise conformally invariant, as the SM is with all masses set to zero, and one might have expected 
this trace to vanish. In (\ref{tranom})
\vspace{-1mm}
\be
E =R_{\a\b\g\l}R^{\a\b\g\l} - 4 R_{\a\b}R^{\a\b}  + R^2\,,\qquad
C^2 = R_{\a\b\g\l}R^{\a\b\g\l} - 2 R_{\a\b}R^{\a\b}  + \sdfrac{1}{3} R^2
\label{ECdef}
\vspace{-1mm}
\ee
are the Euler-Gauss-Bonnet invariant and the square of the Weyl conformal tensor respectively, given in terms of the Riemann curvature 
tensor $R_{\a\b\g\l}$, with $R_{\a\b} = R^\g_{\ \a\g\b}, R= R^\a_{\ \a}$, and the ${\cal L}_i$ denote invariant Lagrangians of any 
gauge fields to which the massless or light QFT degrees of freedom are coupled, such as $\cL_F = F_{\a\b}F^{\a\b}$ for light charged
particles coupled to electromagnetism, or $\cL_G = {\rm tr}\, \{G_{\a\b}G^{\a\b}\}$ for light quarks coupled to the 
$SU(3)^{color}$ gluonic gauge fields of QCD. The $b, b', \b_i$ coeffcients in (\ref{tranom}) are finite dimenionless coeffients 
(in units of $\hbar$) that depend only upon the number and spin of the massless conformal fields contributing to the anomaly, 
analogous to the central term in $D=2$, and are therefore dependent upon the low energy QFT, independently of a UV cutoff 
or completion at the Planck scale. One might also add to (\ref{tranom}) an additional $\sq R$ term with an arbitrary coefficient, 
but this term can be derived from a local $R^2$ action and is therefore better classified with the local Wilsonian EFT terms and 
not as part of the true anomaly.

Like the chiral anomaly in QCD, the conformal anomaly is intrinsically a non-local quantum effect in terms of the original
field (metric and curvature) variables. However, the anomaly effective action can also be recast in the compact local form
\vspace{-1mm}
\be
S_{\!\cA}[g; \vf]=  \sdfrac{b'\!}{2\,}\!\!\int\!\!d^4\!x\sqrt{-g}\,\Big\{\!\!-\!\left(\sq \vf\right)^2  + 2\,\big(R^{\a\b}\! - \!\tfrac{1}{3} Rg^{\a\b}\big)
\,(\pa_\a\vf)\,(\pa_\b\vf)\Big\}  + \sdfrac{1}{2} \!\int\!\!d^4\!x\,\cA\,\vf   
\label{Sanom}
\vspace{-1mm}
\ee
by introducing an additional local scalar field $\vf$. This scalar $\vf$ is closely related to the conformal factor $\s$ of the spacetime metric $g_{\a\b}= e^{2\s}\bar g_{\a\b} $, 
as is clear from the Wess-Zumino (WZ) consistency condition~\cite{AntMazEM:1992, MazEMWeyl:2001, EMSGW:2017}
\vspace{-1mm}
\be
S_{\!\cA}[e^{-2 \s} g; \vf] = S_{\!\cA}[g; \vf + 2 \s] - S_{\!\cA}[g; 2 \s]\,.
\label{SanomWZ}
\ee
satisfied by (\ref{Sanom}). This WZ identity, required by the Weyl cohomology of the anomaly in field configuration 
space~\cite{MazEMWeyl:2001}, stringently constrains the form of the effective action (\ref{Sanom}), forbidding {\it e.g.} 
any simple polynomial terms of the form $\int \!d^4\!x\!\sqrt{-g}\,\vf^n$ that would violate (\ref{SanomWZ}). This identity 
also shows that although $\vf$ is similar in some respects to the dilaton of string theory, it should be distinguished 
from it, as a collective mode composed of a quantum {\it correlated pair} of massless SM fields contributing to 
$\lag \hat T^{\a\b}\rag$, which is not present in classical GR, and associated with explicit breaking by the anomaly 
rather than SSB, motivating the distinct term of {\it conformalon} for $\vf$. 

Since $\vf$ has zero scaling dimension, (\ref{Sanom}) is a technically (marginally) relevant operator of dimension $4$ in the 
sense of Wilson \cite{Wilson:1975}, and hence should be added to the classical E-H action for the low energy EFT of gravity. 
Since (\ref{Sanom}) is the general solution of (\ref{SanomWZ}), up to Weyl invariant terms, it may be regarded as a general 
addition to the EFT, with $b, b'$ coefficients to be determined by experiment, as they may also receive unknown contributions 
from gravity itself.

Variation of (\ref{Sanom}) with respect to $\vf$ gives the linear eq.~of motion
\vspace{-1mm}
\be
\D_4 \vf \equiv  \na_\a\! \left(\na^\a\na^\b +2R^{\a\b} - \tfrac{2}{3} R g^{\a\b} \right)\!\na_{\!\b} \vf
=\sdfrac{1}{2}\! \left( E - \tfrac{2}{3} \sq R\right) + \sdfrac{1}{2b'}\big( b C^2 +  \sumi \,\b_i\cL_i\big)
\label{phieom}
\vspace{-1mm}
\ee
whose general solution is easily found for simple curved backgrounds, while variation of $S_{\!\cA}$
with respect to the metric gives the covariantly conserved stress-energy tensor
\vspace{-1mm}
\be
T_{\!\cA}^{\a\b}[\vf] \equiv \sdfrac{2\!}{\!\!\!\sqrt{-g}} \, \sdfrac{\d}{\d g_{\a\b}} \ S_{\!\cA}[\vf]
\label{Tphi}
\vspace{-1mm}
\ee
whose trace $g_{\a\b}T_{\!\cA}^{\a\b}[\vf] = \big\lag \hat T^\a_{\ \ \a} \big\rag$ is the anomaly (\ref{tranom}).
The detailed form of $T_{\!\cA}^{\a\b}[\vf]$ given in~\cite{EMVau:2006,EMZak:2010,EMSGW:2017}, and has been
evaluated in several fixed background spacetimes, such as the Schwarzschild BH and de Sitter cosmological spacetimes.

Being derived from non-local quantum fluctuations of massless fields, the effect of $S_{\!\cA}$ or $T_{\!\cA}^{\a\b}[\vf] $
and the quantum correlations they encode need not be negligible on the null horizon of a BH or de Sitter space, even
when local curvatures remain small there. In fact (\ref{Tphi}) generally grows without bound like $(r-\rH)^{-2}$ as 
$r\to \rH$ as the horizons of either of these spacetimes are approached, {\it cf.}~eq.~(\ref{Thoriz}) below 
and Refs.~\cite{EMVau:2006,EMZak:2010}. This fourth power diverging behavior follows from the conformal nature of horizons and 
simple dimensional analysis of the stress tensor as a dimension four operator~\cite{MazEMWeyl:2001}, the very same properties of 
the near horizon geometry and stress tensor utilized in the AdS/CFT correspondence~\cite{Witten:1998,Maldacena:1999} and 
Fefferman-Graham expansion~\cite{FefferGra:1995,HenSkend:1998,ImbSchwTheiYank:2000}. The horizon divergences of the 
stress tensor associated with the conformal anomaly effective action are generic, in the sense that a precise fine tuning of the 
integration constants of (\ref{phieom}) for $\vf$, and quantum states of the underlying CFT are necessary to eliminate them, 
which is a set of measure zero. 

Thus the effective action (\ref{Sanom}) of the quantum conformal anomaly can have large {\it macroscopic} effects in the near 
vicinity of BH horizons. Near the horizon even massive fields of the SM may be treated as massless, for $m \ll M_{Pl}$, and
their quantum vacuum polarization and correlation effects described by $\vf$ do not decouple. Thus $S_{\!\cA}$ amounts to 
a well-defined modification of Einstein's GR, fully consistent with, and in fact generally {\it required} by first principles of QFT and 
general covariance, with no additional assumptions, that is a relevant term in both the mathematical sense of Wilson~\cite{MazEMWeyl:2001},
and in the quantum physics of BHs.

\section{The Three-Form Gauge Field Associated with the Anomaly}

Of the several terms in the trace anomaly the Euler-Gauss-Bonnet density $E$ is distinguished by its topological character.
Because it is a total derivative of a coordinate gauge dependent quantity, its integral is a topological invariant, analogous
properties it shares with the more familiar Chern-Simons current and $\ve_{\a\b\m\n} F^{\a\b}F^{\m\n}$ of the axial anomaly.
This topological character of the anomaly is again an infrared feature, on arbitrarily large scales far greater than the
microscopic Planck scale. Since the total derivative
\vspace{-1mm} 
\be
E  =  -\sdfrac{1}{3!\!}\,\ve^{\a\b\m\n} \,  \na_{[\a} A_{\b\m\n]} \equiv -\sdfrac{1}{4!\!}\, \ve^{\a\b\m\n} \, F_{\a\b\m\n}
\label{EJA}
\ee
automatically defines a $3$-form abelian potential $A_{\a\b\g}$, and its corresponding $4$-form field strength $F_{\a\b\m\n}$,
the term in (\ref{Sanom}) linear in $\vf$ and $E $ can then be written as
\be
S_{\rm int} [\vf; A] = \sdfrac{b'\!}{2} \int d^4 x\sqrt{-g}\, E \, \vf = \sdfrac{1}{3!\!} \int d^4 x\sqrt{-g} \, J^{\a\b\g}A_{\a\b\g}
\label{Sint}
\ee
by an integration by parts, where
\vspace{-2mm} 
\be
 J^{\a\b\g} = -\sdfrac{b'\!}{2}\, \ve^{\a\b\g\l} \na_\l \vf
\label{Jphi}
\ee
is the $3$-form abelian current, which is covariantly conserved, $\na_\g  J^{\a\b\g} = 0$ in a general background metric.
The $J\cdot A$ interaction (\ref{Sint}) for this current with the $3-$form gauge potential $A_{\a\b\g}$ is the analog of the 
$J\cdot A$ interaction of point charges with the $1$-form potential of electromagnetism, but now describing a 
$3-$surface charge density of an extended object~\cite{HenTeit:1986}. The worldtube of a BH or cosmological
horizon with topology ${\mathbb R} \times {\mathbb S^2}$ is exactly such a $3$-dimensional extended object
in the vicinity of which the conformalon derivative $\pa_\l \vf$ is large and the current $ J^{\a\b\g}$ is localized. 

\section{EFT of Dynamical Vacuum Energy}
\label{Sec:VacE}

The final ingredient necessary for the construction of the low energy EFT of gravity relevant to vacuum energy  
is the independent observation that a constant $\L$ term in GR is equivalent to introducing a $4-$form gauge field strength
\vspace{-2mm}
\be
F = \sdfrac{1}{\,4!}\, F_{\a\b\g\l}\, dx^\a \wedge dx^\b \wedge dx^\g \wedge dx^\l
\ee
with the `Maxwell' action
\be
S_{\! F} =  -\sdfrac{1}{2\vk^4}\int F \wedge ^*\!\!F=  -\sdfrac{1}{\,48\, \vk^4}\! \int d^4x \sqrt{-g} \  F_{\a\b\g\l}F^{\a\b\g\l}
= \sdfrac{1}{2\vk^4}\!\int d^4x \sqrt{-g}\, \tF^2
\label{Maxw}
\ee
where $^*\!F = \frac{1}{\,4!}\, \ve_{\a\b\g\l}\, F^{\a\b\g\l} \equiv \tF$ is the Hodge dual scalar to $F$. 
This is in analogy to and generalization of QED in two spacetime dimensions for which the classical Maxwell action 
\be
-\frac{1}{4e^2} \int d^2x \sqrt{-g}\, F_{\a\b}F^{\a\b}= \frac{1}{2e^2} \int d^2x \sqrt{-g}\, \tF^2
\label{Max2D}
\ee
where $\tF= \frac{1}{2} \ve_{\a\b}F^{\a\b}= E$ is the electric field in one space dimension. The stress tensor corresponding 
to (\ref{Max2D}) is $g_{\a\b} E^2/2e^2$, which is equivalent to a cosmological term, since $E$ contains no propagating degrees 
of freedom, and is constrained by Gauss' Law to be a spacetime constant in $D\!=\!2$, in the absence of sources. In $D\!=\!2$ the
electric charge $e$ has mass dimension one, while in $D\!=\!4$, $\vk$ also carries mass dimension one if the field strength
tensor $F_{\a\b\g\l}$ has mass dimension $4$.

Now the `Maxwell eq.'~following from variation of (\ref{Maxw}) with respect to $A_{\a\b\g}$ is
\be
\na_\l F^{\a\b\g\l} = 0\qquad {\rm or} \qquad d\, ^*\!F = 0
\label{FnoJ}
\ee
in the absence of sources. Thus (\ref{FnoJ}) implies $\tF \!=\! \tF_0$ is a spacetime constant,
and contains no propagating degrees of freedom. This is a general property of a $p$-form gauge field where $p\!=\!D$
is matched to the number of spacetime dimensions. Moreover in this case the energy-momentum-stress tensor for $F$ is
\be
T^{\a\b}_{\! F} = \frac{2\!\!}{\!\sqrt{-g}} \frac{\d S_{\! F}}{\d g_{\a\b}} = 
-\frac{1}{4!\, \vk^4}\left(\sdfrac{1}{2} g^{\a\b} F^{\g\l\m\n}F_{\g\l\m\n} - 4 F^{\a\l\m\n}F^{\b}_{\ \,\l\m\n}\right)
= -\frac{1} {2\vk^4}\, g^{\a\b} \tF^2
\label{TF}
\ee
which is entirely equivalent to a cosmological term in Einstein's eqs., with the the identification
\be
\L_{\rm eff} = \frac{4\p \GN}{\vk^4}\, \tF^2 
\label{Lameff}
\ee
the effective positive cosmological constant term for $\tF\!=\!\tF_0$ constant. 

The $4$-form field strength and Maxwell action (\ref{Maxw}) for it is entirely equivalent to a positive cosmological
term in Einstein's eqs., and may be freely exchaged for $\L$ by the equivalence (\ref{Lameff}), at the classical level
and in the absence of any sources. Note, however that the integration constant $\tF_0$ is adjustable and can be fixed by a boundary
condition, so that a vanishing cosmological term simply corresponds to the vanishing of the corresponding sourcefree `electric'
field strength $\tF_0$ in infinite empty space, without any fine tuning. Thus this reformulation of the cosmological term in terms of 
$\vk$ and $\tF$ shifts attention away from UV divergences of QFT to macroscopic boundary conditions solving the classical 
constraint eq.~instead.

Now the new element is to identify the $4$-form field strength $F$ of (\ref{Maxw}) with that of the topological 
Euler-Gauss-Bonnet term (\ref{EJA}) of the effective action (\ref{Sanom}) of the conformal anomaly. Then {\it if}
$A$ can be varied {\it independently} of the metric, the $J\cdot A$ interaction (\ref{Sint}) and the derivative of the scalar 
conformalon $\vf$ would provide a source current (\ref{Jphi}) for the $3$-form gauge potential. This independent
variation is possible {\it only} in a general Einstein-Cartan space for fermions minimally coupled to gravity through the
spin connection, {\it cf.} \cite{EMEFT:2022} for details. The $3$-form current is concentrated on an extended tube 
of topology ${\mathbb R} \times {\mathbb S^2}$ if the normal derivative to the tube boundary ${\bf \hat n}\cdot\vec \na \vf$ is
significant there. Then the value of the field strength dual scalar $\tF$ interior to the tube will differ from its exterior value 
by an amount proportional to $\vk^4$ and the change in $\vf$ from interior to exterior of the world tube. Thus the value of the 
effective cosmological term (\ref{Lameff}) as determined by the value of $\tF^2$ will no longer be constant everwhere is space, 
but can change at the tube boundary, if $\vf$ does. This classical $\tF$ or equivalently $\vf$ is the gravitational vacuum condensate,
postulated in~\cite{MazEMPNAS:2004}. 

The low energy EFT of gravity therefore proposed is
\be
S_{\rm eff}[g;\vf;A] = \sdfrac{1}{16\pi G} \!\int\!\! d^4x \sqrt{-g}\, R + S_{\!\cA'}[g;\vf] + S_{\rm int}[\vf;A] + S_{\! F}[A]
\label{Seff}
\ee
where $S_{\!\cA'}$ is given by (\ref{Sanom}) minus the $E- \frac{2}{3}\sq R$ term which has been separated off and taken
account of by (\ref{Sint}) instead. To this effective action for gravity may be added possible higher derivative local terms such as 
$C^2$ and $R^2$ whose effects are known to enter at the UV Planck scale, and the action $S_{\! cl}$ of classical matter or radiation.
Thus, since $S_{\! F}$ by itself is equivalent to a cosmological constant term, the two middle terms of (\ref{Seff}) comng
from the conformal anomaly, together with the identification of the current (\ref{Jphi}) are the proposed additions to the EFT of gravity.

One may remark also that the parameter $\vk$ in $S_{\! F}$ is a kind of topological susceptibility of the gravitational vacuum,
analogous to and a natural generalization of the chiral susceptibility of QCD, associated with instantons and its non-trivial vacuum structure~\cite{Venez:1979,VendiVecc:1980}. Although one might expect the value of $\vk$ to be of order of $M_{Pl}$, there
is no reason {\it a priori} for the two scales to be related, and they are initially distinct, as $\L_{QCD}$ and $f_\p$ are in QCD, 
to be related perhaps in the full UV theory.

\section{Gravitational Condensate Stars}
\label{Sec:GravCond}

The EFT (\ref{Seff}) makes calculations of the boundary layer and interior of gravitational condensate stars 
feasible from first principles for the first time, reducing a nearly intractable quantum backreaction problem to a well-defined 
set of {\it classical} eqs., namely (\ref{phieom}) for $\vf$, that for the frame dependent $3$-form $A_{\a\b\g}$, and the 
classical Einstein eqs.~with the stress tensor sources (\ref{Tphi}) and (\ref{TF}) for the mean geometry. For spherical symmetry
these become ordinary differential eqs.~in $r$ for two metric functions and the radial profiles $\vf (r)$ and $\tF (r)$, which can be 
solved by standard methods. 

One should note that the terms in $T_{\!\cA}^{ab}[\vf]$ following from the anomaly action $S_{\!\cA}$ are proportional 
to $\hbar$, and their relative magnitude to the strictly classical terms is controlled by the dimensionless parameter~\cite{MazEMPNAS:2004,MazEM:2015}
\be
\e\equiv \sqrt{\sdfrac{\hbar c}{G}}\, \frac{1}{M}= \frac{M_{Pl}}{\!M} \simeq 1.1 \times 10^{-38}\ \frac{M_\odot\!}{\!M} \ll 1
\label{eps}
\ee
which is extremely small for BH masses ranging from $M\! \sim\! M_\odot$ for a solar mass BH, or even smaller for
$M\sim 10^6\!\leftrightarrow\!10^9\, M_\odot$ of supermassive BHs in the centers of galaxies. For this reason the quantum 
effects of (\ref{Sanom}) are generally quite negligible, unless greatly enhanced as they are near event horizons.
The physical thickness of the gravastar thin boundary layer may be estimated by finding the value of $r$ at which the spacetime 
curvature produced by the coherent vacuum polarization stress energy density~\cite{EMVau:2006,EMZak:2010} 
\vspace{-2mm}
\be
T^{\,\a}_{\!\!\cA \ \b}[\vf] \sim \frac{\k_{_H}^4}{\left(1- \sdfrac{\rH}{r}\right)^{\!2}}\ {\rm diag}\,(-3,1,1,1) \to \infty
\qquad {\rm as} \qquad r \to \rH
\label{Thoriz}
\ee
becomes comparable to the classical Riemann curvature $\sim \k_{_H}^2$ of Schwarzschild spacetime. 
This occurs at a $\D r\! \simeq\! L_{Pl}$ away the BH classical horizon. Because of the Schwarzschild line element
this corresponds to a physical proper thickness of the boundary layer (gravastar membrane) of 
\be
\ell \simeq \sqrt{L_{Pl}\, \rH} \simeq \sqrt{\e}\, \rH 
\simeq 2.2\,\times\,10^{-14}\!\sqrt{\sdfrac{\! M}{M_\odot}\!} \ \, {\rm cm} \gg L_{Pl}
\label{ellest}
\ee
or the order of the size of an atomic nucleus for $M\!\sim \!M_{\odot}$. Although too small to have yet been detectable
in astrophysical or gravitational wave (GW) observations, this is still many orders of magnitude larger than the extreme microscopic 
Planck scale $L_{Pl}$, so that a mean field EFT treatment of the boundary layer according to (\ref{Seff}) is possible. 

The smallness of $\e$ also makes the method of matched asymptotic expansions, familiar in Prandtl boundary layers of hydrodynamic 
and acoustic flows at large Reynolds' number ({\it e.g.}~airflow around wings and within shock waves) ideally well-suited to the
present problem~\cite{BendOrzag}. Well away from this boundary layer, the energy density due to the quantum anomalous terms 
in $T_{\!\!\cA}^{\a\b}[\vf]$ is ${\cal O}(\e^2)$, and completely negligible. Hence well  outside this boundary layer, classical GR 
applies with $\L \!=\! \L_{\rm eff}$ piecewise constant. In the exterior region $\L_{\rm eff}$ is also ${\cal O}(\e^2)$ and may be taken 
to vanish with high accuracy, so that the classical Schwarzschild BH geometry is recovered. In the interior, again far from the 
boundary, $\L_{\rm eff}\!\simeq\!3/\rH^2\propto 1/M^2$, whose value can adjust itself to the total mass from $\tF^2$, the 
eq.~of state is that of the gravitational vacuum condensate, $p_{_V}\!=\! -\r_{_V}$, according to (\ref{TF}), again to accuracy 
$\e^2$, and the static patch of de Sitter space is recovered as in~\cite{MazEMPNAS:2004,MazEM:2015}.

Within the boundary layer, defined by (\ref{ellest}), $\vf, \tF$ and the metric vary rapidly and the fourth order terms 
in $T_{\!\cA}^{\a\b}[\vf]$ become comparable to the classical Einstein terms. The number of integration constants 
(four) in the fourth order eq.~(\ref{phieom}) assures that the functions $\vf(r),\vf'(r)$ can be matched through the thin membrane 
layer to satisfy the boundary conditions of a finite soln.~at $r\!=\!0$ and asymptotic flatness as $r \to\infty$. 
Obtaining the actual profiles of the functions of $r$  through the boundary layer will require a numerical solution, 
facilitated by the existence of an actional functional (\ref{Seff}) to be extremized.  Carrying this program to
completion would establish the existence of the gravitational condensate solution of the EFT of (\ref{Seff}), 
as well as its most important properties, including its stability to small perturbations, at least in the spherically symmetric case,
and verify the surface tension $\t_{_S}$ found previously in the strictly classical limit $\e \to 0$~\cite{MazEM:2015}. 
As a bonus, the extremal action principle of (\ref{Seff}) will allow the physical interpretation of the solution as a minimum of 
the appropriate zero temperature effective potential, as the size of the object of fixed mass $M$ and the condensate energy 
density of the interior are varied, describing the quantum phase transition boundary layer of a gravastar in the semi-classical 
mean field approximation.

\begin{wrapfigure}[14]{r}{6cm}
\vspace{-7.4mm}
\includegraphics[width=.48\textwidth,viewport=0 -100 600 385,clip]{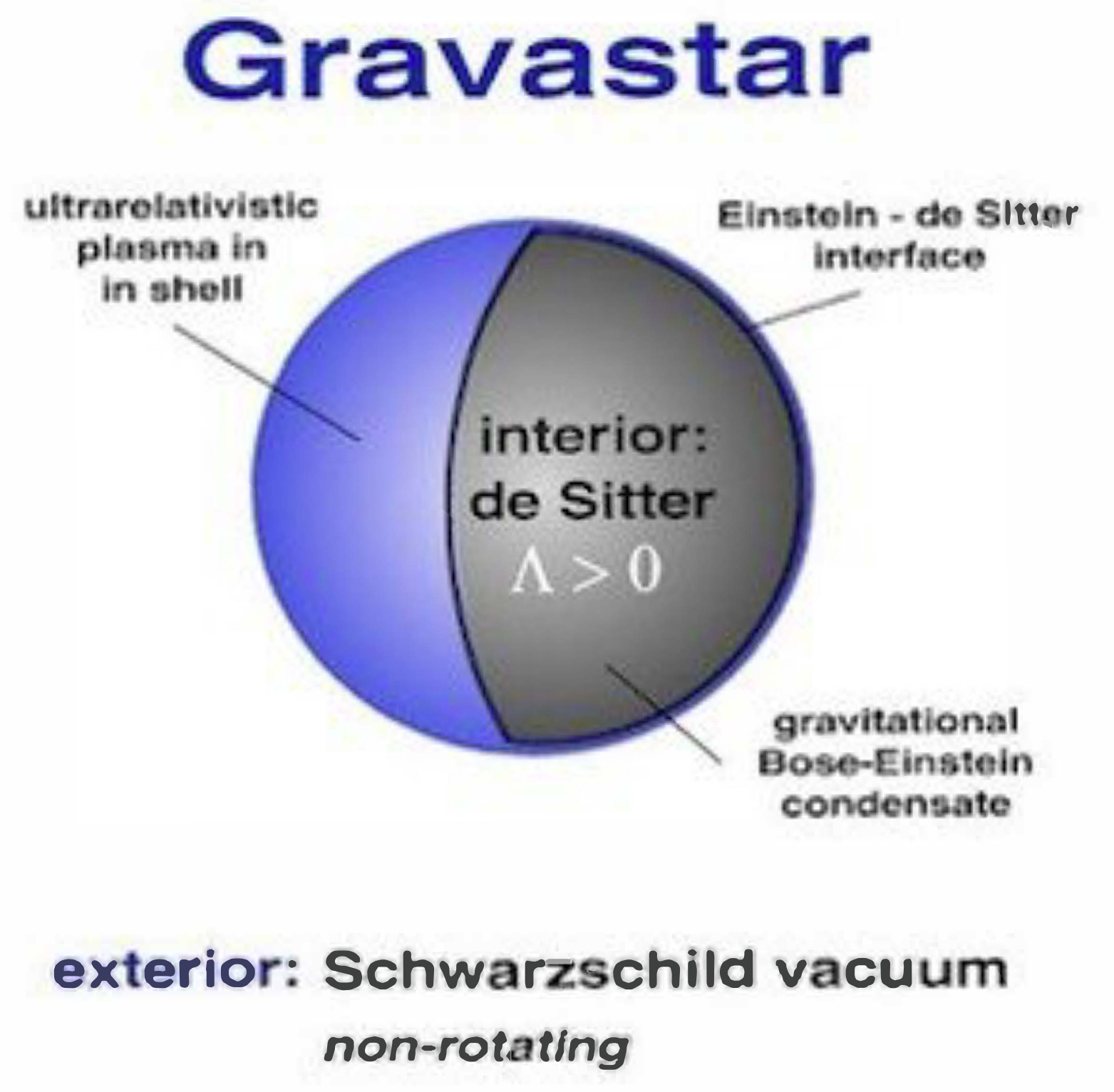} 
\vspace{-2.2cm}
\caption{Schematic representation of a spherically symmetric gravitational condensate star with a de Sitter interior.} 
\label{Fig:Schematic}
\end{wrapfigure}

It should be noted that the smallness of $\e$ in (\ref{eps}) implies a deeply redshifted surface layer, from which 
any radiation emitted can escape to infinity to be observed only if emitted within a pinhole solid angle of order $\e$ around 
the normal to the surface. Thus an object with this degree of compactness would appear to a distant observer just as 
dark as a BH, even if the surface could thermalize the energy of infalling matter and re-radiate all of it thermally rather 
than absorbing it. The authors of \cite{CarbLibVis:2018} show that observations of accretion but lack of substantial 
emission from Sag A* can yield a bound, $\e \lesssim 10^{-17}$ at best, impressive, but still well short of the estimate 
(\ref{eps}), and argue that the thermalization and re-emission assumptions of~\cite{BrodNary:2007,BrodLoebNary:2009} 
are likely to fail as well. Hence the astrophysics of an isolated gravastar will be essentially the same as that of a BH: cold, dark, 
and compact. This is the reason that the observations so far cannot rule out the existence of such a very thin boundary layer 
at a distance (\ref{ellest}) above $\rH$.

\section{Cosmic Acceleration and Large Scale Structure with Dynamical Dark Energy}
\label{Sec:DDE}

With the EFT of the conformal anomaly (\ref{Sanom}) coupled to the $3$-form potential and $4$-form abelian field strength term 
described in Sec.~\ref{Sec:VacE}, we are now in a position to propose for the first time a detailed study of Dynamical Dark 
Energy (DDE) in cosmology based on fundamental theory, derive its predictions, and rigorously test those predictions 
against obervations and Large Scale Structure. The value of the cosmological dark energy in this EFT is related to
the cosmological horizon scale by boundary conditions of the Universe in the large, rather than UV cutoffs.

Assuming the SM, the only free parameter in the EFT (\ref{Seff}) is the constant $\vk$ controlling the coupling of the $\vf$ gradient 
to the $3$-form gauge field and vacuum energy. Note from the form of the anomaly $\cA$ in (\ref{tranom}), any deviation from exact 
homegeneity and isotropy will lead in general to $F_{\m\n}F^{\m\n} \neq 0$ for the photon radiation field, and 
${\rm tr}\, \{G_{\m\n}G^{\m\n}\} \neq 0$ for the electroweak and QCD color gauge fields in the unconfined phase of the early 
Universe. This will induce changes in the conformalon field $\vf$ through its eq.~of motion (\ref{phieom}), which will then cause the 
field strength $F_{\a\b\m\n}$ and hence the vacuum energy $F_{\a\b\m\n}F^{\a\b\m\n}$ to change. After the transition to the 
confining phase of QCD, baryonic matter will still contain non-vanishing gluonic condensates and thus still act as a source for 
$\vf$, thereby coupling non-relativistic baryonic matter to dynamical vacuum energy as well. Thus although $\vf$ is {\it not} 
an inflaton, it is a scalar that is well-grounded in QFT of the SM and can produce backreaction effects on the vacuum energy 
when fluctuations away from exact homogeneity and isotropy are admitted. It permits interaction between both radiation and 
matter with DDE, in which adiabaticity of the matter and radiation components will no longer be satisfied in general, in effect 
introducing a bulk viscosity into the cosmological fluid.  If $\L_{\rm eff} \!\propto\! F_{\a\b\m\n}F^{\a\b\m\n}$ does not remain 
constant in the de Sitter phase deviations from the $\L$CDM cosmology are to be expected. 

\section{Scalar Gravitational Waves in the EFT}

\begin{wrapfigure}[17]{r}{7.2cm}
\vspace{-5mm}
\includegraphics[width=.4\textwidth,viewport=0 -100 640 550,clip]{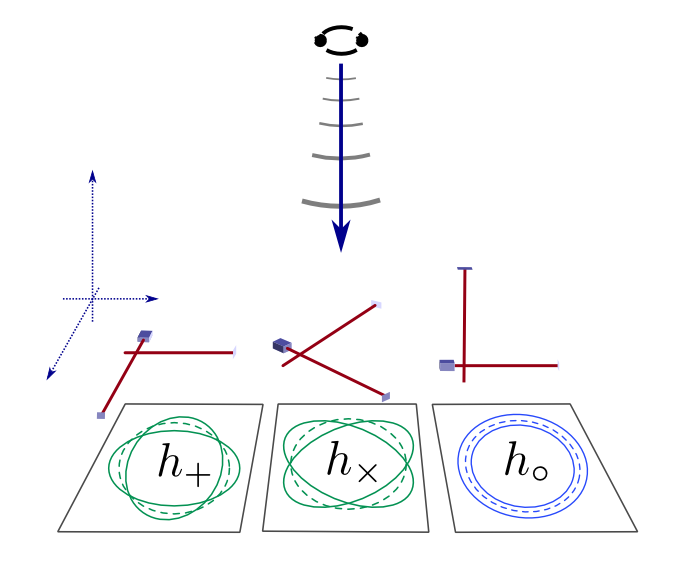} 
\vspace{-1.3cm}
\caption{Two tensor polarization modes, $h_+$ and $h_\times$, and the scalar mode $h_0$. 
The optimal orientation of the GW detector arms for each mode is shown
for the case when the GW is propagating directly from above.} 
\label{Fig:SGWs}
\end{wrapfigure}

Einstein’s classical GR predicts just two transverse, traceless spin-$2$ GW polarizations denoted $h_+$ and $h_\times$. The limited 
GW data so far is consistent with this~\cite{LIGO:2019}. A most striking consequence of (\ref{tranom}) and (\ref{Tphi}) coupled to 
GR in the EFT of (\ref{Seff}) is the prediction of an additional long range dynamical scalar gravitational field $\vf$, which 
couples to the spin-$0$ conformal factor of the metric, and turns it into a {\it propagating} degree of freedom. In Einstein's classical 
GR the spin-0 sector of metric deformations $g_{\a\b} \to e^{2\s} g_{\a\b}$ is constrained, and does not propagate, analogously 
to the Coulomb potential of classical electromagnetism. However in the semi-classical Einstein equations with (\ref{Tphi}) as source, 
when linearized around flat space new wave solutions are obtained, which are scalar gravitational waves (SGWs), {\it i.e.} a spin-0
‘breather’ GW polarization in addition to the two transverse, traceless polarizations of the classical Einstein theory~\cite{EMSGW:2017}.
The analysis of~\cite{EMSGW:2017} also shows that although (\ref{phieom}) is a fourth order differential equation, the linearized
propagating SGWs of the spacetime metric obey a conventional {\it second order} wave equation as a result of the diffeomorphism
constraints. They carry {\it positive} energy, thus evading the problems afflicting modifications of GR involving higher order local metric derivatives, found previously even in linear perturbations of flat space~\cite{Stelle:1977}. 

The amplitude of the conformalon scalar $\vf$ is determined by the QCD gluonic condensate in high density nuclear matter through
 (\ref{phieom}) and is related therefore to the eq.~of state in neutron star (NS) cores. As a result, SGWs are expected to be produced in NS-NS and NS-BH mergers, as well as in the early universe from the quark-gluon plasma. The magnitude of the SGW at frequency 
$\w$ and distance $r$ from a NS merger can be calculated in the EFT to be~\cite{EMSGW:2017}
\be
h_0 = - \frac{G}{3r}\, e^{-i\w ( t-r) } \, \int d^3 \bx \, \exp (-i \w \br \cdot \bx)\,\cA_\w (\bx)  \simeq 0.5 \times 10^{-21} \left( \frac{100 \, Mpc}{r}\right)
\label{SGW}
\ee
where $\cA_\w$, the Fourier transform of the trace anomaly, given by the gluonic condensate in high density QCD, estimated to be 
of order $250$ to $500$ MeV/fm$^3$. Thus the SGW amplitude is of order $10^{-21}$ for a NS merger event like GW170817 observed 
at a distance of 40 Mpc, at the sensitivity bounds of current GW detectors and above the increased sensitivity limits of upgraded detectors. 

The novel scalar polarization $h_0$ can be disentangled from the two transverse polarizations $h_+$ and $h_\times$, only by 
triangulation with three or more GW detectors, or with a clear EM multi-messenger counterpart, lifting the degeneracy with the 
source direction, as illustrated in Fig.~1. When four and eventually five GW detectors are fully operating at their design sensitivities, 
the GW signals detected in the several dozen merger events per year expected at these increased sensitivities will allow precision tests 
of GR and the SGW polarization predicted from the conformal anomaly effective action (\ref{Sanom}). Needless to say, any positive
detection of SGW's in NS mergers would provide evidence in favor of the EFT (\ref{Seff}), and the QFT modification of Einstein's GR it entails.

\section{Summary and Outlook}
\label{Sec:Sum}

A first principles EFT approach to including quantum effects in gravity has been developed, based upon 
the conformal anomaly of the energy-momentum tensor of QFT, the effective action corresponding 
to it, and the long range massless scalar degree of freedom this effective action implies. This leads to a well-defined modification 
of classical GR, fully consistent with, and in fact required by quantum theory, the Standard Model, and the Equivalence Principle.
In this White Paper submitted as part of the Snowmass 2022 process, this EFT is proposed to address the outstanding
problems at the intersection of QFT and Einstein's GR, as holding the promise of resolving both the BH information paradox 
and accounting for the value of the cosmological vacuum dark energy driving the accelerated expansion of the Universe. 
The proposal relates the two problems at different scales to the concept of the vacuum as a kind of gravitational Bose-Einstein 
condensate, whose value can change depending upon external boundary conditions. 

In gravitation theory the demonstration of a stable non-singular endpoint of gravitational collapse consistent with quantum 
theory would resolve the longstanding information paradox besetting classical BHs, and result in a reevaluation of the importance 
of coherent quantum effects in macroscopic self-gravitating systems. Gravitational condensate stars would then become 
viable candidates for gravitationally fully collapsed compact objects in the Universe, and the EFT would provide a predictive theoretical
framework for distinguishing them from classical BHs by future GW and multi-messenger observations anticipated in this decade
and the next. In astrophysics this would stimulate a systematic search for these objects by their GW and 
electromagnetic signatures. In fundamental HE physics these developments could point to entirely new ways of bringing 
gravity within the  framework of a consistent quantum theory, reverberating to the foundations of both.  

There are a number of open problems which need to be addressed in order to fulfill this program. 
Among them (but by no means a complete list) are:
\begin{itemize}[labelsep=8pt,leftmargin=20pt]
\vspace{-2mm}\item  Demonstration of Existence and Stability of Spherical Gravitational Condensate Stars 
\vspace{-2mm}\item  Extension to Axial Symmetry and Rotating Gravastars with Angular Momentum 
\vspace{-2mm}\item  Interaction of SM Fields with the Gravastar Surface for Astrophysical Observations
\vspace{-2mm}\item  Investigation of Gravitational Wave and Electromagnetic 'Echo' Signals from the Interior
\vspace{-2mm}\item  Diffeomorphism Constraints \& Identification of Remaining Ghost-Free Degrees of Freedom
\vspace{-2mm}\item  Canonical Space + Time Splitting \& Cauchy Initial Value Formulation of the EFT
\vspace{-2mm}\item  Time Dependent Evolution and Gravitational Collapse in the EFT
\vspace{-2mm}\item  Detailed Emission of Scalar Gravitational Waves in Radial Collapse and NS Mergers
\vspace{-2mm}\item  New Cosmological Models of Dynamical Dark Energy in the Early and Present Universe
\vspace{-2mm}\item  Comparison with Numerical Simulations and Obsevational Data
\end{itemize}
\vspace{-1mm}

This is clearly a large program that would involve a wide variety of researchers and expertise, from purely theoretical
investigations, to numerical relativity simulations, to analysis of large amunts of data. It therefore seems timely to present this 
proposed EFT to the HE and wider physics community in the Snowmass process for further critical development, theoretical 
consistency checks, and eventual confirmation or refutation by observations in this decade. The promise of discovery of the 
first deviations from Einstein's classical GR due to quantum physics could revolutionize our understanding of both BHs and the 
the quantum vacuum, and point to new directions for the eventual synthesis of gravitation and quantum theory.

\bibliographystyle{myapsrev4-2}
\bibliography{gravity.bib}

\end{document}